\documentclass[epj]{webofc}
\usepackage[varg]{txfonts}   
\woctitle{XLVI International Symposium on Multiparticle Dynamics}
\begin{document}
\title{Fractal aspects of hadrons}
%
%

\author{Airton Deppman\inst{1}\fnsep\thanks{\email adeppman@gmail.com} \and
        Eugenio Meg\'ias\inst{2,3}\fnsep\thanks{\email{emegias@mppmu.mpg.de}}
}

\institute{Instituto de F\'isica da Universidade de S\~ao Paulo, Rua do Mat\~ao, 187 - Travessa R, S\~ao Paulo, Brazil
\and
           Max-Planck-Institut f\"ur Physik (Werner-Heisenberg-Institut), F\"ohringer Ring 6, D-80805, Munich, Germany
\and
          Departamento de F\'{\i}sica Te\'orica, Universidad del Pa\'{\i}s Vasco UPV/EHU, Apartado 644,  48080 Bilbao, Spain
          }

\abstract{%
  The non extensive aspects of $p_T$ distributions obtained in high energy collisions are discussed in relation to possible fractal structure in hadrons, in the sense of the thermofractal structure recently introduced. The evidences of self-similarity in both theoretical and experimental works in High Energy and in Hadron Physics are discussed, to show that the idea of fractal structure of hadrons and fireballs have being under discussion for decades. The non extensive self-consistent thermodynamics and the thermofractal structure allow one to connect non extensivity to intermittence and possibly to parton distribution functions in a single theoretical framework.
}
\maketitle
\section{Introduction}

In the 50's the existing experimental data on High Energy Collisions (HIC) evidenced the formation of a system in thermodynamic equilibrium. Indeed, Fermi proposed a thermodynamic model to describe hadronic collisions at high energies~\cite{Fermi}. The main issue then was to explain how a short-lived system (half-life of a few $fm/c$)
could reach equilibrium so fast. This system was called {\it fireball}.
A few years later Hagedorn~\cite{Hagedorn65} proposed his Self-Consistent Thermodynamics (SCT) theory that was able to explain several features of hot hadronic systems. His starting point was a rather weird definition for fireballs, stating that

{\it A fireball is (*) a statistical equilibrium (hadronic black-body radiation) of an undetermined number of all kinds of fireballs, each o which, in turn, is considered to be (goto *)``}

\noindent The fact that the word fireball appears in the definition of fireball itself shows that this is a kind of recursive definition, not common in Physics. Self-reference is known to arise logical problems in many fields~\cite{Goedel}, but Hagedorn was able to use this definition to build the entire thermodynamics description of fireballs.

The SCT gave predictions for some quantities that could be easily accessed experimentally. It predicted the transverse momentum distribution of the particles produced in the decay of the fireball, given by 
\begin{equation}
 \frac{dN}{dp_T}\bigg|_{y=0}=gV \frac{p_T m_T}{(2\pi)^2}\exp\left( \frac{m_T}{T} \right) \,,
\end{equation}
where $g$ is a constant, $V$ is the volume of the system, $m_T=\sqrt{p_T^2+m^2}$.
Another prediction was the hadron mass spectrum, given by
\begin{equation}
   \rho(m)=\frac{\gamma}{m^{5/2}} \exp(\beta_o m) \,, \label{Hagmspec}
\end{equation}
where $\beta_o=1/T_o$ and $T_o$ is a parameter of the theory.
One interesting aspect of the theory was the prediction of a limiting temperature for the fireball, known today as Hagedorn temperature, $T_H$, which is numerically equal to the parameter $T_o$, i.e., $T_H=T_o$.
The comparison between theory and experiment brought a sudden recognition of the importance of Hagedorn's theory, and its impact in HEP is notorious. Frautschi proposed a similar theory based exclusively in hadron structure, stating that {\it ''hadrons are made of hadrons``}, again using self-reference~\cite{Frautschi}. With this so-called bootstrap model he obtained the same hadron mass spectrum formula shown in Eq.~(\ref{Hagmspec}).

The success of SCT prompted the development of ideal gas models for hadronic systems, the Hadron Resonance Gas (HGR) models~\cite{Hagedorn_HRM, Agasian, Tawfik, Megias_HRM, Huovinen, Borsanyi, Bazavov, Arriola}, that were able to explain many features of high energy collisions. Before that, an important paper by Dashen, Ma and Bernstein~\cite{DashenMaBernstein} had explained, through Dyson-Schwinger expansion, how a strongly interacting system could behave under some conditions as an ideal gas of resonant particles. Another important consequence of SCT came with the advent of the quark structure of hadrons, that was used by Cabibbo and Parisi~\cite{CabibboParisi} to propose that Hagedorn temperature was not a limiting temperature but a transition temperature between the confined quark and the deconfined quark regimes of hadronic matter. The last phase is known as Quark-Gluon Plasma and is one of the most interesting issues in nowadays Nuclear Physics. Despite its initial success and its important consequences, with the results coming from accelerators able to deliver particles at higher energies than those from existing at that time, it was found that SCT was not able to explain correctly the outcome of the new HEP data. Indeed, Hagedorn himself proposed, in substitution to this thermodynamics theory, an empirical model based on QCD~\cite{Hagedorn83}. Since his theory gave so many correct information about hadronic systems, the question imposes itself is: what went wrong with Hagedorn's theory?

In this work a possible answer to the question formulated above is proposed. It is shown that hadrons can present, in a very specific form, a fractal structure. Then it is shown that such fractals must be necessarily described through the Tsallis statistics~\cite{Tsallis1988}, a generalization of the Boltzmann-Gibbs-Shannon statistics. After this, a generalization of the self-consistent theory developed by Hagedorn accommodating the generalized statistics is presented, what can be called Non Extensive Self-Consistent Thermodynamics (NESCT). Comparison of the results of NESCT and experimental data, some of its consequences and applications are discussed.

\section{Thermofractals}

At the same time that Hagedorn was developing his SCT theory, Mandelbrot was proposing the existence of systems with fractionary dimensions, the fractals~\cite{Mandelbrot}. Fractals are systems that present some special characteristics as scaling symmetry and self-similarity. Today the fractal structure is known to be present in many complex systems~\cite{Tel,Schroeder}. It is interesting to note that Tsallis motivation to propose his generalized entropy was fractal systems~\cite{Tsallis1988}.

The self-similarity is an evident feature of fireballs and hadrons according to their definitions by Hagedorn and Frautschi, respectively, as described above. However many other evidences point to the self-similar structure of hadrons: fractal dimension has been identified through the analysis of intermittence in experimental distributions obtained in HEP experiments~\cite{Hwa, HwaPan, Bialas_Peschanski, Bialas_Peschanski2,DreminHwa,Hegyi1,Hegyi2, Sarkisyan, Sarkisyan2, Kittel, Wolf, Bozek}; a Parton Distribution Function (PDF) that describes the proton structure based on fractal properties was shown to fit experimental data rather well~\cite{Lastovicka, Dremin_Levtchenko}; direct evidences of self-similarity in experimental data has been observed~\cite{WWselfsymmetry, Tokarev, Beck}.
In the following a system where fractal structure appears in its thermodynamical characteristics rather than in its geometrical aspects is introduced.

Thermofractals are systems presenting the following properties~\cite{Thermofractal}:
\begin{enumerate}
 \item The total energy is given by $U=F+E$, 
where $F$ corresponds to the kinetic energy of $N'$ constituent subsystems and $E$ corresponds to the internal energy of those subsystems, which behaves as particles with an internal structure.

\item The constituent particles are thermofractals. The ratio $\langle E \rangle /\langle F \rangle$ is constant for all the subsystems. However the ratio $E/F$ can vary according to a distribution which is self-similar (self-affine), $\tilde{P}(E)$, that is, at different levels of the subsystem hierarchy the distribution of the internal energy are equal (proportional) to those in the other levels. 

\item At some level $n$ in the hierarchy of subsystems the phase space is so narrow that one can consider $\tilde{P}(E_n) d\,E_n=\rho dE_n$,
with $\rho$ being independent of the energy $E_n$.
\end{enumerate}

It can be shown that its thermodynamical potential~\cite{Thermofractal} can be written as
\begin{equation}
 \Omega= \int_0^{\infty} A \bigg[1+\frac{\varepsilon}{NkT}\bigg]^{3N/2} \tilde{P}(\varepsilon) d\varepsilon \,, \label{Oeps}
\end{equation}
where $N=N'+2/3$.
Since $\Omega= \int_0^{\infty} P(U) dU$
and $P(U)$ must be related to $\tilde{P}(\varepsilon)$ because of the second property of thermofractals, Eq.~(\ref{Oeps}) must be satisfied simultaneously with the following identity:
\begin{equation}
 P(U)=\tilde{P}(\varepsilon)\,, \label{similarity}
\end{equation}
expressing the self-similarity of thermofractals.

The simultaneous solution for Eqs.~(\ref{Oeps}) and~(\ref{similarity}) is obtained with
\begin{equation}
 P(\varepsilon)= A \bigg[1+\frac{\varepsilon}{NkT}\bigg]^{-\frac{3N}{2}\frac{1}{1-\nu}}\, \label{selfsimilar}\,.
\end{equation}

Introducing the index $q$ by $q-1=(2/3N) (1-\nu)$ and the effective temperature $\tau=2(1-\nu)T/3$
one finally obtains
\begin{equation}
 P(\varepsilon)=A e_q(-\varepsilon/k \tau)\,, \label{Tsallis_weight}
\end{equation}
where
\begin{equation}
e_q(x)=\bigg[1+(1-q)x \bigg]^{\frac{1}{1-q}}
\end{equation}
is the Tsallis q-exponential factor characteristic of the nonextensive statistics. It replaces the Boltzmann exponential factor by generalizing it, since for $q \rightarrow 1$, $\tilde{P}(\varepsilon)$ becomes the well-known exponential factor of Statistical Mechanics.

This result is a possible answer to the question proposed in the Introduction: what went wrong with Hagedorn's theory? Today we know that Hagedorn's fireball is a fractal, or more specifically a thermofractal. The result above shows that thermofractals must be described by Tsallis statistics, not by Boltzmann statistics as Hagedorn did.
The next questions are: is it possible to generalize Hagedorn's theory with Tsallis statistics? And could the generalized theory describe experimental data? In the next section these questions will be answered.

\section{Non Extensive Self-Consistent Thermodynamics}

NESCT is a generalization of the SCT theory by imposing the self-consistency principle from Hagedorn in the non extensive statistics from Tsallis~\cite{Deppman}. The basic ingredients are the two forms of partition function for fireballs proposed by Hagedorn, only this time using the Tsallis q-exponential factor in Eq.~(\ref{Tsallis_weight}), that is,
\begin{equation}
 Z_q(V_o,T)=\int_0^{\infty}\sigma(E)[1+(q-1)\beta E]^{-\frac{q}{(q-1)}} dE  \,,
 \label{Zq1}
\end{equation}
where $\sigma(E)$ is the density of state of the fireballs as a function of its energy, and 
 \begin{equation}
 \begin{split}
 \ln[1+Z_q(V_o,T)]= & \frac{V_o}{2\pi^2}\sum_{n=1}^{\infty}\frac{1}{n}\int_0^{\infty}dm \int_0^{\infty}dp \, p^2 \rho(n;m) [1+(q-1)\beta \sqrt{p^2+m^2}]^{-\frac{nq}{(q-1)}} \,,
\label{Zq2}
\end{split}
\end{equation}
which is the partition function for an ideal gas of fermions and bosons with mass $m$. The sum in $n$ results from the expansion of the logarithm function, and $\rho(n;m)=\rho_f - (-1)^n \rho_b$ is related to the fermionic and bosonic mass spectra. In the case  $n=1$, which will be used here, $\rho(1;m)=\rho_f + \rho_b=\rho(m)$ is the hadron mass spectrum.

Since fireballs are thermofractals and present self-similarity, the following weak-constraint must hold
\begin{equation}
 \ln[\rho(m)]=\ln[\sigma(E)]\,.
\end{equation}
The main task now is to find functions $\rho(m)$ and $\sigma(E)$ that satisfy the equality above and simultaneously let the two forms of partition function identical to each other. This can be achieved by choosing~\cite{Deppman}
\begin{equation}
   \rho(m)=\frac{\gamma}{m^{5/2}}[1+(q_o-1) \beta _o m]^{\frac{1}{q_o -1}} \qquad \textrm{and} \qquad\sigma(E)=bE^a\big[1+(q_o-1)\beta _o E\big]^{\frac{1}{q_o -1}} \,. \label{qmasspectrum}
\end{equation}
With those functions both forms of partition function result in
\begin{equation}
Z_q(V_o,T) \rightarrow b \Gamma(a+1)\left( \frac{1}{\beta-\beta _o}\right)^{a+1} 
\label{asymptotic_Zq}
\end{equation}
with $a+1=\alpha=\gamma V_o/2\pi^2 \beta^{3/2}$. From here one can see that there is a singularity at $\beta=\beta_o$, and therefore the limiting temperature is found.

Therefore it is possible to generalize Hagedorn's self-consistency principle within Tsallis statistics and as a result one gets a new hadron mass spectrum formula, given by $\rho(m)$ in Eq.~(\ref{qmasspectrum}), the limiting temperature resulting from the singularity in Eq.~(\ref{asymptotic_Zq}), now expressed in terms of the Tsallis temperature, but also a universal entropic index, $q$, which is characteristics of any hadron. It is worthy to mention that both $T_o$ and $q_o$ are parameters in the hadron mass spectrum, therefore they are related to the hadronic structure. As in the case of the Hagedorn's theory, here also the way to test experimentally the predictions of the theory is through the transversal momentum distribution, given by
\begin{equation}
 \frac{dN}{dp_T}\bigg|_{y=0}=gV \frac{p_T m_T}{(2\pi)^2} e_q\left(\frac{m_T}{T} \right) \,.
\end{equation}

The predictions of NESCT have been compared with HEP experimental data for $p_T$ distributions, showing a good agreement between calculation and data~\cite{Lucas, Cleyman_Worku, Sena, AzmiCleymans1, AzmiCleymans2}, resulting in $q_o=1.14$ and $T_o=62$~MeV. Also, the hadron mass spectrum formula has been used to describe the known hadronic states, resulting again in a very good agreement with data and leading to values of $q_o$ and $T_o$ very similar to those obtained with $p_T$ analysis~\cite{Lucas}. The extension to finite chemical potential was performed in~\cite{Megias}, and was applied to the analysis of neutron star stability~\cite{Debora}.
A comparison between the results from NESCT and LQCD data also has shown a good agreement between both calculations~\cite{DeppmanLQCD}, in this case without any adjustable parameter. From the thermofractal structure one can obtain the fractal dimension of hadrons, what was done in Ref.~\cite{Thermofractal}, resulting in $D=0.69$, a value that is close to that resulting from intermittence analysis, around $D=0.65$~\cite{Ajienko, Rasool, Singh, Albajar,Ghosh}

These analysis confirms that Hagedorn's self-consistency principle holds for hot hadronic systems even at energies as high as those found at LHC, when the appropriate statistics is used. The Tsallis statistics, on the other hand, must be used in the construction of fireballs analysis because it takes into account the complex partonic structure of hadrons. The fact that the fractal dimension obtained from the $q_o$ and $T_o$ values obtained in the $p_T$ analysis or in the mass spectrum analysis shows that the hypothesis that hadrons presents a fractal structure as that described for thermofractals is a reasonable one. This is evidenced by experimental data, as explained in the introduction, and by the analysis of intermittency.

\section{Conclusions}

In this work it is discussed the fractal aspects of fireballs and hadrons as defined respectively by Hagedorn and Frautschi. By the introduction of the concept of thermofractal, a system presenting fractal structure in its Thermodynamic functions, it is shown that such systems must be described by Tsallis statistics rather than Boltzmann statistics.

The self-consistency principle proposed by Hagedorn is then generalized assuming Tsallis statistics, resulting in a non extensive statistics that predicts a new hadron mass spectrum formula, a limiting temperature and a universal entropic index for hadrons. This result shows that it is possible to generalize Hagedorn's theory to obtain power-law spectra for the particles produced in high energy collisions.

Several works have shown that the power-law spectra resulting from the non extensive statistics can correctly describe the general properties of particle production in HEP experiments, confirming the existence of the limiting temperature and of the universal entropic index. Also, in the present work it is discussed that the new hadron mass spectrum describes the observed hadronic states better than Hagedorn's formula. Calculations from Latice QCD are also in good agreement with the generalized thermodynamics model.

Finally, it is shown that the fractal dimension of the thermofractal obtained with values for temperature and entropic index that result from analysis of $p_T$ distributions in high energy collisions are in agreement with those found in analyses of intermittence in experimental data from high energy collisions.

\begin{acknowledgement}
This work was supported by the Brazilian agency, CNPq, under grant 305639/2010-2, by Plan Nacional de Altas Energ\'{\i}as Spanish MINECO grant FPA2015-64041-C2-1-P, and by the Spanish Consolider Ingenio 2010 Programme CPAN (CSD2007-00042). E.M. would like to thank the Instituto de F\'{\i}sica of the Universidade de S\~ao Paulo, Brazil, for their hospitality and support during the completion of the initial stages of this work. The research of E.M. is supported by the European Union under a Marie Curie Intra-European fellowship (FP7-PEOPLE-2013-IEF) with project number PIEF-GA-2013-623006, and by the Universidad del Pa\'{\i}s Vasco UPV/EHU, Bilbao, Spain, as a Visiting Professor.
\end{acknowledgement}

\end{document}